\documentclass[twocolumn]{revtex4}
\usepackage{latexsym}
\usepackage{amsmath}
\usepackage{times}
\usepackage{amssymb}
\usepackage{fancyheadings}
\usepackage[T1]{fontenc}
\usepackage[utf8]{inputenc}
\pdfoutput=1
\usepackage{url}
\usepackage{hyperref}
\usepackage{color}
\usepackage{graphicx}
 \usepackage{epsfig}

\begin{document}

\title{\bf Self-Organised Criticality and Emergent Hyperbolic
  Networks---Blueprint for Complexity  in Social Dynamics}
\author{Bosiljka Tadi\'c$^{a,b}$}
\affiliation{$^a$Department of Theoretical Physics, Jo\v zef Stefan Institute,
Jamova 39, Ljubljana, Slovenia; $^b$Complexity Science Hub Vienna,
1080 Wien, Austria}
\begin{abstract}
\noindent
Online social dynamics based on human endeavours exhibit prominent complexity in the emergence of new features embodied in the appearance of collective social values. The vast amount of empirical data collected at various websites provides a unique opportunity to quantitative study od the underlying social dynamics in full analogy with complex systems in the physics laboratory. Here, we briefly describe the extent of these analogies and indicate the methods from other science disciplines that the physics theory can incorporate to provide the adequate description of human entities and principles of their self-organisation. We demonstrate the approach on two examples using the empirical data regarding the knowledge creation processes in online chats and questions-and-answers. Precisely, we describe the self-organised criticality as the acting mechanisms in the social knowledge-sharing dynamics and demonstrate the emergence of the hyperbolic geometry of the co-evolving networks that underlie these stochastic processes.\\
\end{abstract}

\maketitle
\thispagestyle{fancy}

\section{Introduction\label{sec-intro}}
One of the most apparent features of complexity is the emergence of a
new functional property of the ensemble of interacting units
which does not exist at any of its constituents. 
In this regard, complex systems have multi-scale dynamics and exhibit
collective phenomena that mostly resemble self-organized criticality
in physical systems
\cite{SOC-book2013,SOC-reviewPhysRep2014,Avalanches-FunctMatGeophys2017,SOC-biology2011,SOC-neuronalNat2007,SOC-Brain-review2014}.
The majority of physical and biological systems with these
properties link their dynamics with complex topological
characteristics in the coordinate or phase space
\cite{SOC-Brain-review2014,we-chargetransport,topol-function1-4,TT2004,dorogovtsev-criticality}. Therefore, one of the 
fundamental questions of the science of complexity regard the role of topology in the emergence of new properties.
 
Mapping the dynamics of a complex system onto a graph (network) is the
first step to address this problem. In this context, a suitable
mapping results in a mathematical graph with nodes and connections
that contain some essential attributes of the original system and its
dynamics. On the other hand, it provides an objective analysis using
graph theory \cite{BB-book,SD-book}. The naturally evolving structures
obey a particular kind of optimisation principle at all stages of
growth. Therefore, the architecture of a graph receive interpretations
regarding the fundamental dynamics and interactions between the
constitutive units of the system. A striking example of this principle
is the community structure in the protein interaction
\cite{PINets-communities} and gene expression \cite{we-genes} networks that can be detected by graph theory methods \cite{community-str}.

Recently, the study of graphs representing various complex systems has
been extended beyond the standard graph-theoretic metric. The use of
methods of algebraic topology \cite{AT1,AT2} enabled revealing the higher
organised structures and related hidden geometries that can appear in
the graph. More specifically, the Q-analysis based on the algebraic topology of graphs identifies elementary geometric shapes or simplexes
(triangles, tetrahedrons and higher order cliques) and how they
connect to each other making the more massive structures, simplicial
complexes \cite{Q1,Q2,Q3}. A particular composition of these basic
geometric shapes can lead to emergent hyperbolicity or a negative
curvature \cite{HB1,HB2}, a measure of the nodes proximity in the metric graph space, which often associates with improved collective
dynamics in natural and technological networks \cite{HBa1,HBa2,HBa3}.

\begin{figure}[!htb]
\begin{tabular}{cc} 
\resizebox{18pc}{!}{\includegraphics{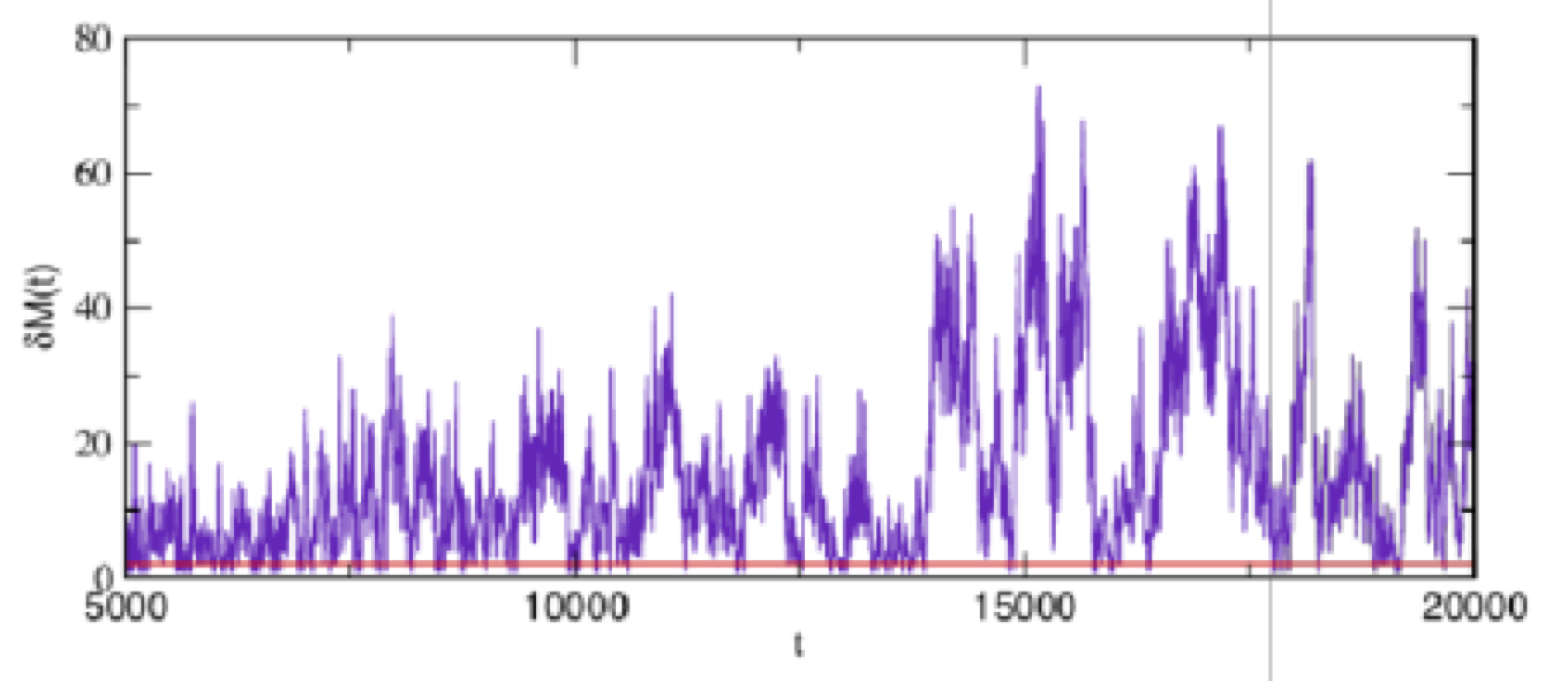}}\\
\end{tabular}
\caption{The noisy response of a driven system in the laboratory contains information on how the number of elementary events contributes to the collective response. The example is Barkhausen noise simulated in the model of the random-field ferromagnet slowly driven by the external magnetic field along the hysteresis loop. Temporal clustering of events---avalanches, can be recognised as the bursts of the signal above the red baseline. For a particular avalanche, the duration $T$ is the distance between the two consecutive intersections of the baseline with the signal while the enclosed area  under the signal between these two points gives the avalanche size $s$. In a self-organised critical state, the avalanche sizes and durations exhibit power-law distributions \cite{BT_PhysA1999}.}
\label{fig-TS-example}
\end{figure}

Among complex systems that have recently driven much attention of the
science community are different types of networks, which embody human
social experience \cite{falk-basset} from patterns of brain activity of participating individuals during social communications
\cite{we-socialBrain,we-brainHB} to large-scale online social graphs
\cite{we-entropy,mitrovic2010a,mitrovic2010b,mitrovic2011,we-MySpace11,we-emoRobotsCMP2014}. These studies are primarily enabled by the vast amount of empirical data obtained by brain imaging in the laboratory \cite{brain-platform} as
well as from online social interaction sites
\cite{social_manifesto,social-book} where users leave information
about themselves and their actions that  are continuously being stored
at the server.
In contrast to the abovementioned networks representing dynamical systems in the physical world, the principles governing social interactions are more complex and depend on various human attitudes during the communication process as well as the contents  (cognitive, emotional) contained in the exchanged messages. Nevertheless, the high-resolution data on social websites contain sufficient information to extract relevant dynamical quantities, which can be analysed by approaches based on the formal analogy to the physical systems in the laboratory. Precisely, the response of a physical system to the external driving force can be measured in the laboratory and often represents a noisy signal (time series of a relevant dynamical variable), see an example in Fig.\ref{fig-TS-example}, from which the avalanching dynamics can be studied. Extracting the time-series of the relevant quantities from the empirical data then
enables the use of time-series analysis, entropy measures, information divergence and self-organised criticality (see \cite{we-entropy,we-SciRep2015} and references there).

In social dynamics, the sequence of events is stored in the empirical data and often contains much other valuable information: id of each user, the action on a particular artefact, the contents of its action---i.e., as a new question, answer, or comment and often to whom the answer was intended (in the case of direct communication between the users).  This additional information can be used to extract the features of human subjects in the process and to describe the nature of the interaction between them, which is governed by a different principle compared with the energy of the interaction in the case of physical particles. Furthermore, with the help of machine-learning methods \cite{ML},
additional information characterising human interactions on the Web
can be extracted, for example, emotional  and cognitive contents of
the exchanged messages \cite{mitrovic2010b,garas-Chats,we-Chats-PhysA,we-SciRep2015}, which make the set of
quantities even richer 
as compared to the physical systems, and also allow the appropriate
theoretical approaches using the agent-based modeling \cite{BT-abm,we_ABM_blognets,we-Chats-exp}.

In this work, we focus on the social dynamics based on
knowledge-sharing communications; we consider the appropriate data from two social sites: \textit{Ubuntu} chats and
questions\& answers site \textit{StackExchange/Mathematics}.  A
detailed analysis of these datasets
\cite{garas-Chats,we-Chats-PhysA,we-SciRep2015} revealed different aspects of the social process of knowledge creation; consequently, these datasets lead to different types of networks \cite{we-Chats-PhysA,we-Chatnets,we-SciRep2015,we-KCNets_PLOS2016}, as it will be described below. Here, we demonstrate that these networks
possess higher topological properties leading to the emergent hyperbolic geometry. Moreover, we show the signatures of the
self-organised criticality in the underlying stochastic processes and
discuss the connections to the hyperbolicity of the networks
co-evolving with these processes.

\section{Emergence of social networks via social knowledge-sharing processes\label{sec-mapping}}
\begin{widetext}
\begin{figure*}[!]
\begin{tabular}{cc} 
\resizebox{22pc}{!}{\includegraphics{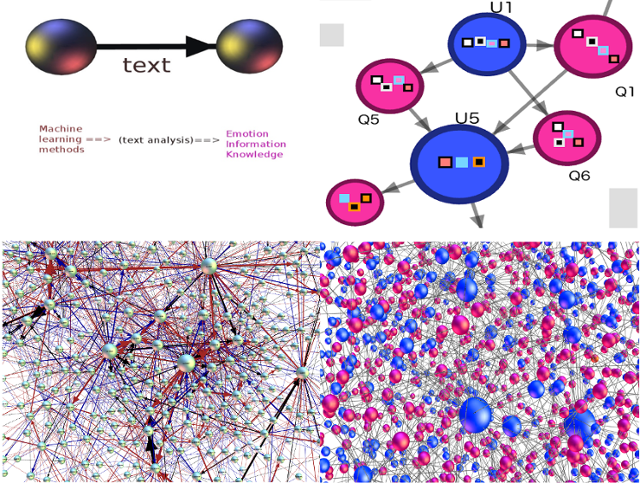}}\\
\end{tabular}
\caption{  (left) Top panel: Schematic representation of User-to-User interaction via the exchange of messages, whose contents can be extracted by Machine-learning methods and attributed to the link between the Users. Bottom: A  close-up of the  network of Users
  (nodes) connected via emotion-carrying links extracted from data of chats in IRC Ubuntu channel; red positive, black negative,
  blue--neutral emotion messages [1].
(right) Top panel: Schematic representation of the indirect interaction between Users (blue) mediated by Questions (red) nodes, which contain the tagged contents of the exchanged information. Bottom: A close-up of the bipartite networks of users (blue) and questions (red nodes) emerging from the sequence of events in Q\&A site StackExchange Mathematics. In the data, the contents of each Question is tagged according to standard Mathematics Classification Scheme.
}
\label{fig-KCP-mapping}
\end{figure*}
\end{widetext}

The social communications can create new knowledge through so-called \textit{meaningful} interactions, which social psychology defines as communications intended to meet the needs of others \cite{kimmerle,book_epistemology}.
Modern technology greatly facilitates these processes by enabling easy access and fast communication between the participants and preserving the data. At the same time, it transforms the system into the online social dynamics, which is still not well understood.  Different websites created with the purpose of knowledge-sharing have their own action rules, which can influence the course of the process and its outcome. Here, we illustrate such differences in two types of
knowledge-sharing systems: Internet-Relay-Chat (IRC) Ubuntu and Questions \& Answers (Q\&A) site Mathematics from StackExchange.  The corresponding empirical
data are mapped onto the network as  illustrated in Fig.\
\ref{fig-KCP-mapping}. 
In particular, depending on the objectives of the study,
the considered contents of the messages exchanged among users can be associated either to the links between
the nodes or another type of nodes introduced.

\begin{figure}[!htb]
\begin{tabular}{cc} 
\resizebox{16pc}{!}{\includegraphics{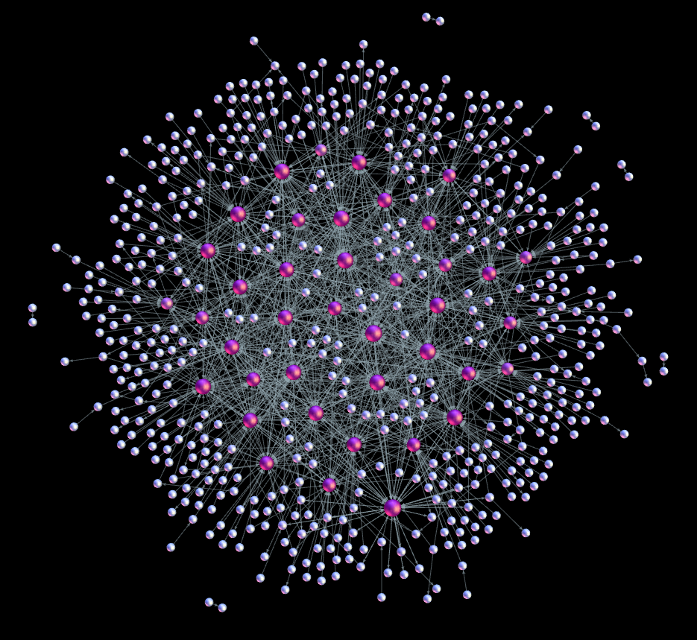}}\\
\end{tabular}
\caption{The emergent network of online chats with Bot and moderators, comprising the central core, and users   (data for this image are from \cite{we-Chatnets}).}
\label{fig-Chats-core}
\end{figure}

\textit{IRC channel Ubuntu} provides
support for the users of the operating system (bringing the
original tribe Ubuntu spirit---“humanity to others” to the cyber
world) \cite{ubuntu-http}. A user places his request on the
channel; then noticed by one of the currently present users, the
problem is either answered by that user or forwarded to one of the
more knowledgeable users (moderators) or the Bot. The Bot has some
predefined answers to many standard questions. About 40 users
voluntarily play the role of moderators. Within the Cyberemotions project \cite{Cyb-http}, a broad set of data from Ubuntu chats were collected together with the text of each message and analysed in works \cite{garas-Chats,we-Chats-PhysA,we-Chatnets,we-Chats-exp,we-Chats-conf,we-Chats-Bot}. The
analysis revealed several interesting observations. Firstly, due to the fast evolution of the events, about ten events per minute, the
attention is focused on the most recent questions, and rarely a search for previous answers occurs.
Consequently, the gained knowledge concerns the involved individuals and not a commonly shared collective experience. Secondly, while some users quickly disappear after getting a satisfactory help, the majority remained active on the channel for very long time (counted in years), keeping in contact with the acquired connections among moderators and continue helping new users. In this way, a stable social network forms through these chatting relations  \cite{we-Chats-conf,we-Chatnets}. It consists of a central core, where the moderators and Bot are mutually interconnected, and a corona made by other users that are connected to them and among themselves, see Fig.\ \ref{fig-Chats-core}. Screening of the text of messages for emotion contents \cite{garas-Chats,we-Chats-exp}
revealed that the most of the links exchanged among the users contain emotional words carrying positive or negative emotion valence; these contents are associated with the connections between users, as shown in Fig.\ \ref{fig-KCP-mapping}.
Furthermore, analysing the emotional arousal, the
degree of the excitement, in the messages exchanged along different
links, we have found \cite{we-Chats-PhysA} that the high-arousal links efficiently keep
the network together. More specifically, its giant cluster appears via a percolation-type transition when the arousals are exceeding a threshold value. We have shown that the emergent structure in
Ubuntu chats data as well as the similar networks obtained through the agent-based simulations of chats \cite{we-Chatnets,we-Chats-Bot} fulfil the test of ‘social hypothesis’, thus representing a right social graph.  For this reason,
it was not surprising that the emotional Bot, designed in the experimental conditions \cite{we-Chats-exp}, was able to impact the emotional state of the users. Performing large-scale simulations of chats with emotional Bots \cite{we-Chats-Bot,we-emoRobotsCMP2014}, we have shown how the Bots manage to polarise the collective emotional states of agents, splitting the networks into two layers with positive and negative emotion flow. The hyperbolicity of the structure of these layers is analysed in Sec.\ \ref{sec:HB}, see Fig.\ \ref{fig-Chats-HB}.

In the case of social knowledge-creation via Questions-and-Answers, the data from the StackExchange Mathematics site were considered \cite{we-SciRep2015,we-KCNets_PLOS2016,we-SOC2017}. Here, each user has its unique ID, and its actions are stored in the open log file. In contrast to chat channels, the process is much slower, approximately ten questions per hour, and evolves in a self-organised manner without any central authority. Also, the users often search for previously answered items when their cognitive content is related to a currently active issue. By its first appearance, each question is tagged by up to 5 tags according to the
standard Mathematics Classification Scheme, for example, Graph theory, Differential geometry, Number theory, Abstract algebra, Probability, Statistics, and so on. In total, over one thousand different tags appear in
all questions in the considered period of four years.   Each posted
question and answer has a unique ID and content as well as the time of publishing and the user who posted or referred to it. Hence, the
interaction between users is effectively going through the questions,
or better to say, through the contents of these questions. For these
reasons, we present the sequence of events as a bipartite graph, see
Fig.\ \ref{fig-KCP-mapping}, where the user-nodes make one partition while the
questions-and-answers are the nodes of the other partition. It is
important to notice that in such a graph no direct link can occur
between the nodes of the same partition, which corresponds to the nature
of the interactions on this site.

Another prominent feature of this process is that the cognitive
contents of each question are systematically observed. Namely, only the users with the expertise in the area mentioned by tags in that question can offer a meaningful answer. In the data, the user’s skill is not explicitly
known but can be derived statistically from the history of its
actions, as we have done in \cite{we-SciRep2015}. In designing the agent-based model of
these processes, see \cite{we-SciRep2015,we-SOC2017}, we create the agents with given expertise
(expressed by a combination of tags) and assume that strict matching
of the agent’s knowledge with the contents of the answered question should apply at least in one tag. The comparisons of the prominent
features of the stochastic processes in the simulated and empirical data confirmed \cite{we-SOC2017} that, indeed, the expertise matching the question contents plays the crucial role in the creation of knowledge by the expansion
of innovation; see precise definition and details in \cite{we-SciRep2015}. Thus, the users
bring new knowledge that accumulates via answers on a particular question. The sequence of events is then mapped onto a growing bipartite network with two types of nodes representing users and artefacts. An example is shown in Fig.\ \ref{fig-QA-bip} corresponds to the subset of the empirical data
where each question-node contains the tag Linear algebra among other tags.  

The tag-matching between the user’s expertise and the contents of the
answered question is illustrated in Fig.\ \ref{fig-Tagnets}. For the objectives of
this work, it is important to notice that the tag-matching constraint
during the social process of questions-and-answers leads to a
non-random appearance of different tags. Namely, being a part of the
expertise of each user, various combinations of tags occur in a
logical connection with each other. Consequently, the network of these knowledge contents emerges reflecting the ways that they were used in the process. In the context of knowledge creation processes, these
networks contain what is called \textit{explicit knowledge}, that is a form of
the collective knowledge used by the participants, and remains for
others to learn from it \cite{kimmerle}. In \cite{we-KCNets_PLOS2016} we have extracted such networks
of tags over consecutive year-long periods and analysed their
hierarchical architecture by algebraic topology methods. An example is shown in Fig.\ \ref{fig-Tagnets} corresponding to the contents used
during the first year of the users' activity.
Moreover, we have shown that the attachment of new tags to the network occurs through so-called innovation channel, which contains new tags added within a specified
short period and the tags to which they attach through the underlying users’ actions. The analysis in \cite{we-KCNets_PLOS2016} revealed that the attachment of
new contents observes the logical structure of mathematics knowledge. This
structure confirms that the network growth principle obeys the
tag-matching constraints at each stage. Here, in Sec.\ \ref{sec:HB} we analyse the
hyperbolic geometry of these tag networks as well as of their innovation channels
appearing over time.

\section{ Hyperbolic geometry of networks emerging in knowledge--sharing social endeavours\label{sec:HB}}
Spontaneous evolution of networks in physical world obeys certain
optimisation principles at different stages of the network growth. The
developed complex architecture often exhibit hidden geometries with
emergent hyperbolicity, the proximity of nodes in the graph-metric space,
which facilitates flow between them.  In social systems, on the other
hand, the co-evolving networks embody human interactions, which are
governed by different principles and depend on communicated cognitive
or emotional contents. Above, we have described how two types of social networks
emerge from knowledge-sharing social endeavours: the graph that
contains explicit knowledge derived from the empirical data of
knowledge-creation processes by Questions and Answers,   and a
social graph built through IRC chats devoted to Ubuntu problems
solving. Here, by testing the Gromov 4-point criterion of hyperbolicity \cite{gromov-4p}, we
show that the structure of these empirical networks is
$\delta$-hyperbolic. 

A generalization of the Gromov notion of hyperbolicity \cite{gromov-4p} has been
applied to networks \cite{HBa1,HBa2,HBa3}; it uses the graph metric space
(shortest paths) to measure  the distance between nodes. Here, we
apply 4-point Gromov delta-hyperbolicity \cite{gromov-4p} test to the adjacency
matrices of the graphs that we described above. Specifically, for an
arbitrary set of four nodes $(A,B,C,D)$ one defines three combinations
of the sums of the pairwise distances $d(A,B)$,  $d(C,D)$, $d(A,C)$, $d(B,D)$,
$d(A,D)$, and $d(B,C)$. Then these combinations are ordered from the
largest to smallest ${\cal{L>M>S}}$, where, for example, the largest sum is ${\cal{L}}=
d(A,B)+ d(B,C)$, middle ${\cal{M}}=d(A,C) + d(B,D)$, and  the smallest
${\cal{S}}=d(A,B)+d(C,D)$. For a $\delta$-hyperbolic graph, there is a small fixed
value $\delta$ such that any four nodes of the graph satisfy the condition
\begin{equation}
2\delta(A,B,C,D) = {\cal{L}} - {\cal{M}} < \delta \; .
\label{eq:hb-condition}
\end{equation}
According to the triangle equality, the upper bound of the difference
between the two largest sums, $({\cal{L}}-{\cal{M}})/2$, is given by the minimal distance $d_{min}$ in the smallest sum, that is  
\begin{equation}
d_{min}\equiv min\{d(A,B),d(C,D)\}  \; 
\label{eq:hb-dmin}
\end{equation}
in the given example.  Therefore, plotting $\delta (A,B,C,D)$ against
the corresponding $d_{min}$ for all 4-tuples of the graph indicates the
possible $\delta$ value if the curve that is indicated by averaging
over all points $\langle \delta \rangle$ saturates for larger distances. A sharp saturation value suggests fixed hyperbolicity. In contrast,
the graph exhibits weak or no hyperbolicity if the corresponding curve
grows sub-linearly or linearly with the distance \cite{HB1,HB2}.  For a given
network, which is defined by its adjacency matrix, we first compute
the matrix of all distances by computing the shortest-paths tree from
each node. The distance matrix is then used to determine the pair
distances and abovementioned their combinations for each randomly
selected 4-tuple of nodes. A large number of 4-tuples is sampled; the
sample-average $\langle \delta \rangle$ corresponding to a given
$d_{min}$ is plotted against $d_{min}$. Additionally, we also track
the largest value  $\delta_{max}$ that occurs
in the whole network and plot it against $d_{min}$.

\begin{figure}[!htb]
\begin{tabular}{cc} 
\resizebox{16pc}{!}{\includegraphics{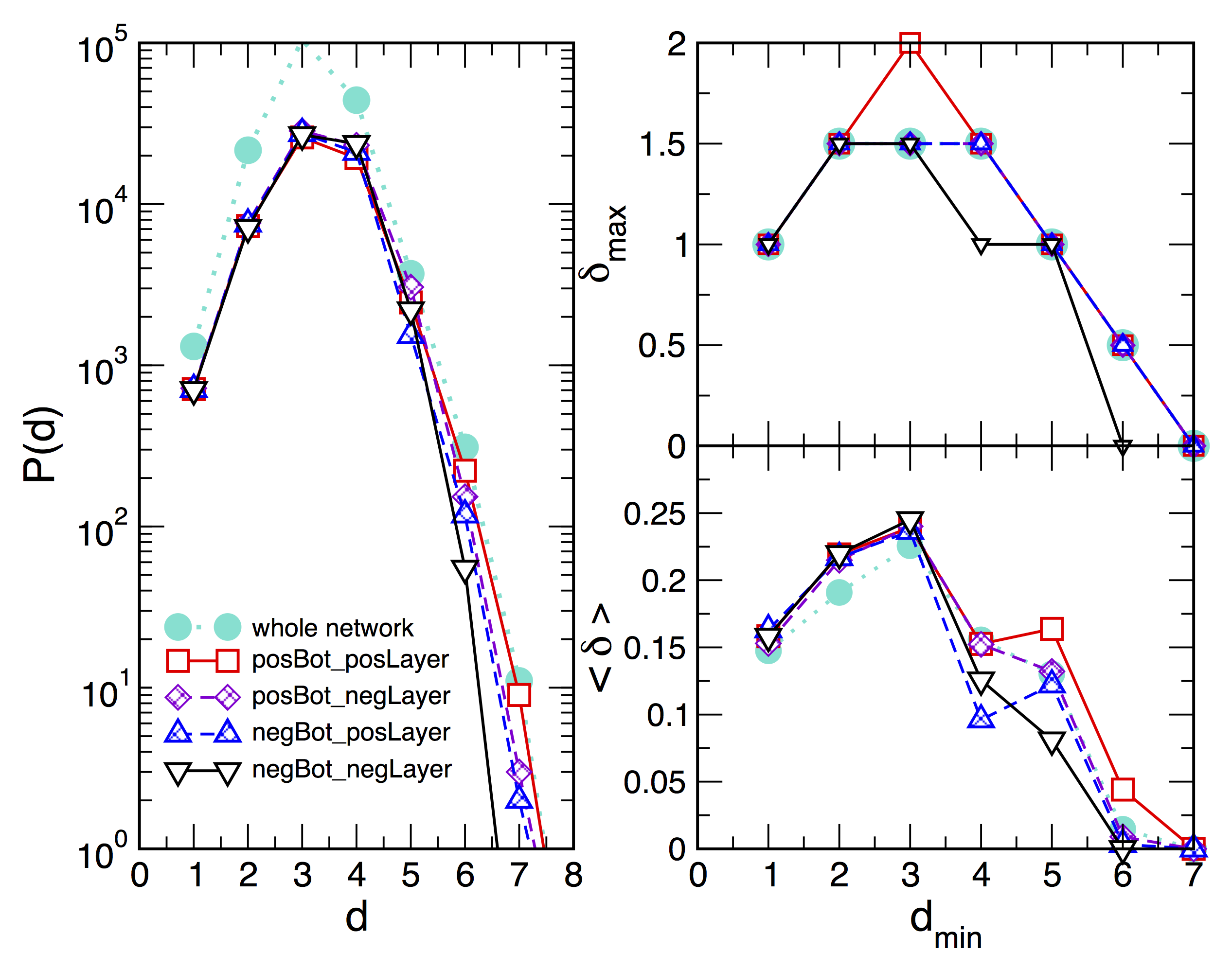}}\\
\end{tabular}
\caption{Histogram of the shortest path distances $d$ between all pairs of nodes $P(d)$, shown in the left panel, and  the average
  (lower panel, right), and maximum hyperbolicity  (upper panel, right) for the chat network and its emotion-flow layers in the presence of Bots, indicated in the legend.}
\label{fig-Chats-HB}
\end{figure}

In the following, we examine $\delta$-hyperbolicity of two types of networks derived from the knowledge-sharing dynamics in online chats channel and questions-and-answers site.
More specifically,  in the online chats simulations in the presence of emotional Bots have been performed \cite{we-Chats-Bot}, resulting in
emotion polarisation of the communicated messages following the
emotion valence which is preferred by the Bot. We analyse the whole
network, see Fig.\ \ref{fig-Chatnet}, and two layers of it which
contain nodes and their connections along which negative/positive
emotion messages flow. The situations corresponding to the Bots
preferring positive (posBot) and negative (negBot) are considered,
see more details in \cite{we-Chats-Bot}. The results are shown in
Fig.\ \ref{fig-Chats-HB} indicating that this social network, as well as its emotion-carrying layers, are $\delta$-hyperbolic with $\delta_{max} \leq 2$.

\begin{figure}[!htb]
\begin{tabular}{cc} 
\resizebox{16pc}{!}{\includegraphics{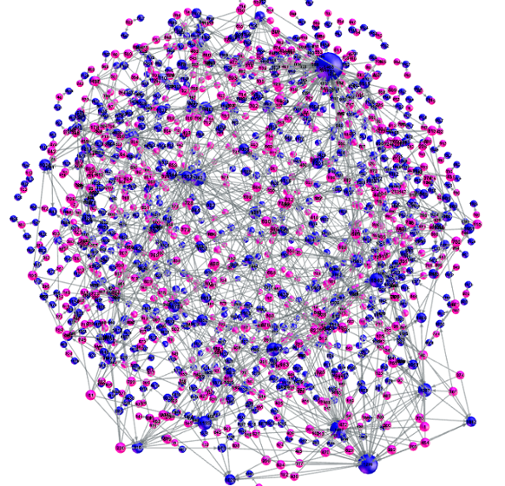}}\\
\end{tabular}
\caption{The bipartite network of users (blue nodes) and questions-and-answers (red nodes) that contain ``Linear algebra'' among other tags. Data for this image are from \cite{we-SciRep2015}. }
\label{fig-QA-bip}
\end{figure}

Although the knowledge of the participating individuals plays a vital
role in building the chat networks, the direct user-to-user contacts
contribute to developing its social-graph characteristics. In Q\&A data,
however, the knowledge contents appear at different levels resulting in two types of networking. First, the users can groups around a particular set of tags, representing their primary interest. An example of content-based bipartite community is shown in Fig.\ \ref{fig-QA-bip}.

\begin{figure}[!htb]
\begin{tabular}{cc} 
\resizebox{16pc}{!}{\includegraphics{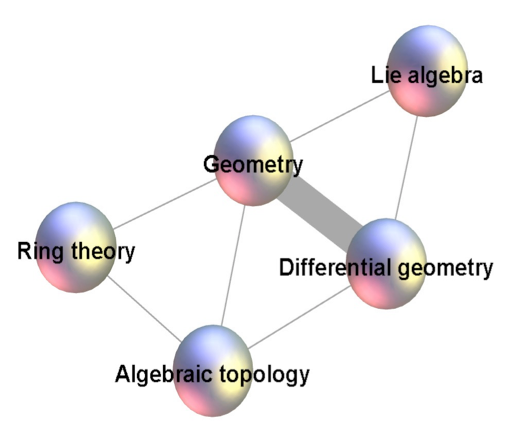}}\\
\resizebox{16pc}{!}{\includegraphics{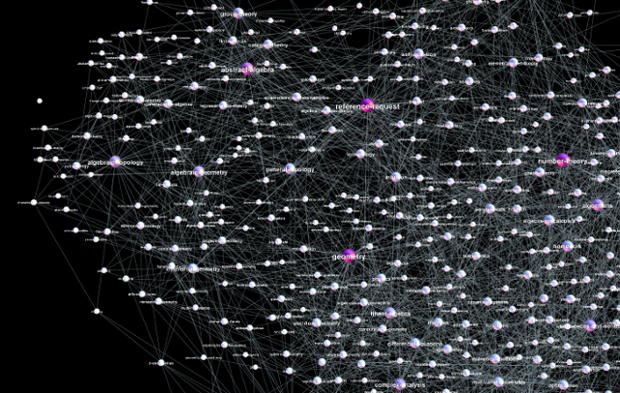}}\\
\end{tabular}
\caption{ Top: Illustration of networking among knowledge contents (tags)  used in the sequence of questions and answers by the users U1 and U5
  in Fig.\ \ref{fig-KCP-mapping}. Bottom: The network of tags emerging
  after the first year from the opening of the site Mathematics of StackExchange. Following the filtering of redundant links  \cite{we-KCNets_PLOS2016}, the network embodies the logical structure of knowledge. }
\label{fig-Tagnets}
\end{figure}
The use of particular knowledge contents (tags) can be systematically
analysed in the considered data and agent-based modelling \cite{we-SciRep2015}, assuming
that the individual knowledge of each user can be tagged
accordingly. For example, Fig.\ \ref{fig-Tagnets} reveals the knowledge contents that
were utilised during the communications by the users U1 and U6,
featuring in Fig.\ \ref{fig-KCP-mapping}. For instance, the four tags (Geometry, Differential
geometry, Algebraic topology, Ring theory) appear to be the user’s  U1 expertise, while three tags (Geometry, Differential geometry, Lie algebra) represent the knowledge of the user U5. Then the interaction via
questions (answers) as in Fig.\ \ref{fig-Tagnets} leads to mutual linking of these tags. 

\begin{figure}[!htb]
\begin{tabular}{cc} 
\resizebox{16pc}{!}{\includegraphics{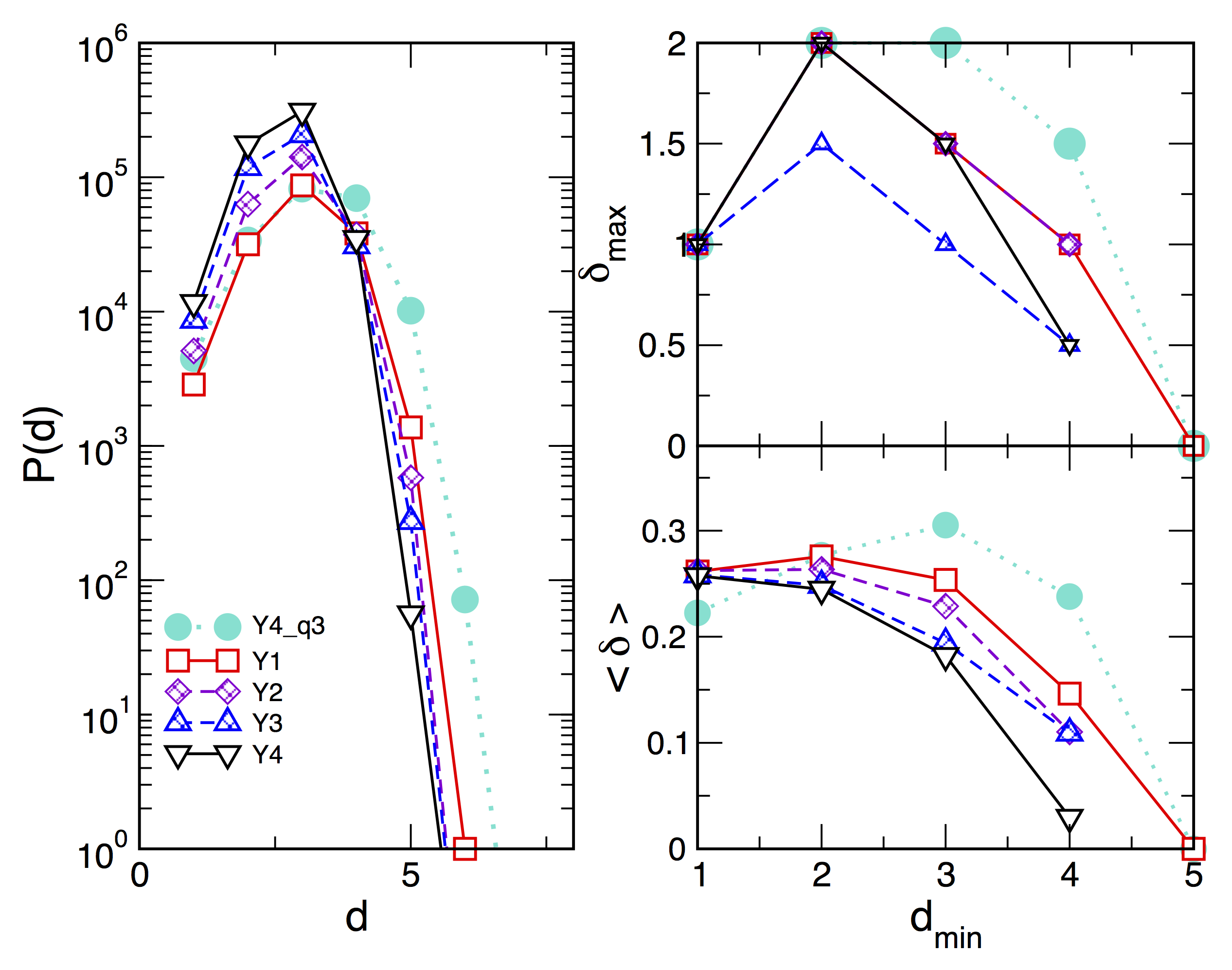}}\\
\end{tabular}
\caption{Histogram of the shortest path distances $P(d)$, left panel, and  the average  (lower panel, right) and maximum hyperbolicity
  (upper panel, right) for the innovation channels of the tag-networks over the years Y1 to Y4.}
\label{fig-Tagnets-HB}
\end{figure}
Next, we divide the whole set of events into the sequence of the specified period (one year). For each period, we obtain the network of tags, which reflects the way that these knowledge contents were used by the participants in all questions and answers occurring within that period. As it was shown in \cite{we-KCNets_PLOS2016}, filtering of these networks is needed to remove spurious connections (precisely defined by null-model) and obtain the network of tags with a given confidence level. The resulting networks, see an example in Fig.\ \ref{fig-Tagnets}, contain
\textit{explicit knowledge} built during this social process.  Then the new tags added to it within a short time at the beginning of the
next period are considered. Together with the previous tags to which they directly link, the newly added contents constitute an \textit{innovation channel} of the knowledge network, for more details, see \cite{we-KCNets_PLOS2016}.

In Fig.\ \ref{fig-Tagnets-HB} we show the results of
hyperbolicity analysis of innovation channels of tags networks over three consecutive year-long periods.
The results show that the all analysed tag-networks appear to have emergent hyperbolic structure.  
The origin of the hyperbolicity can be sought in the appearance of
simplicial complexes of elementary geometric forms (cliques of
different orders); they associate together to form higher organised complexes, which are determined in \cite{we-KCNets_PLOS2016}. As the networks of tags grow
over the years by the addition of innovation channels, their structure becomes increasingly more complex as expressed concerning topology structure vectors \cite{we-KCNets_PLOS2016}. Notably, the network
after the fourth year (Y4) has a shorter diameter and gradually
smaller $\langle \delta\rangle $ while $\delta_{max}$ remains limited below
two. By cutting the Y4-network along the topology level $q=3$, i.e.,
removing the cliques of the orders lower than the indicated $q$, the
higher-order simplexes that appear in the network remain loosely
connected along faces of the order $q$ and larger. Consequently, the
hyperbolic structure changes such that $\delta_{max}   =2$ appears at longer distances, and overall $\langle \delta \rangle $ is larger compared to the whole network. In-depth understanding the links between the mesoscopic structure of the network and its
hyperbolic geometry is currently receiving much attention in theoretical investigations \cite {math-HB-AT}, and can have many practical consequences.

\section{Features of Self-Organised Criticality in Social Dynamics\label{sec:SOC}}
In this regard, by considering the timestamp of the events, we construct different time series of the number of events occurring at a given small time interval as well as the number of events in a small time interval carrying a specified content (cognitive or emotional). Then, performing the analysis of these time series, we can extract the features of the underlying stochastic processes.  Here, the assumption
is that the time series contains "the breath” of the dynamical social
system, in full analogy to laboratory measurements, for instance,
measured Barkhausen noise in the field-driven random ferromagnetic
materials, given in Fig.\ \ref{fig-TS-example}.  Then the analysis
of these time series is performed to test the key features of the
self-organised criticality \cite{SOC-book2013}. In particular, 
we identify the avalanches of events and their potential
scale-invariance, the temporal correlations, and fractal features of
the collective fluctuations
\cite{we-entropy,we-SciRep2015,we-SOC2017}, see below.

\begin{figure}[!htb]
\begin{tabular}{cc} 
\resizebox{16pc}{!}{\includegraphics{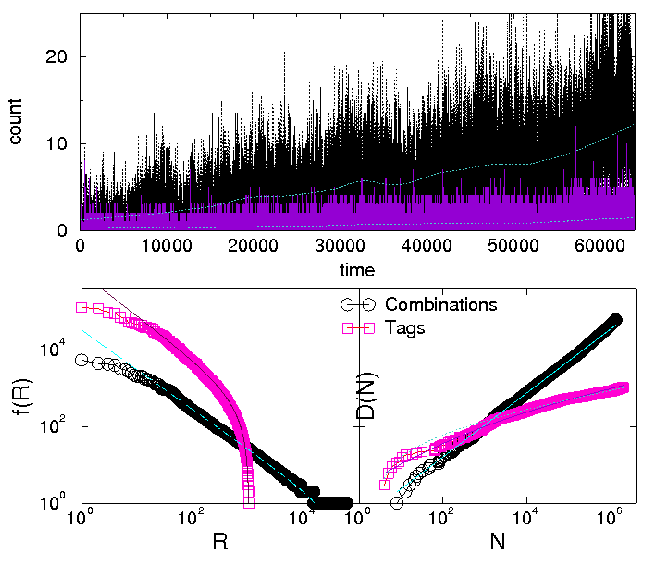}}\\
\end{tabular}
\caption{  From Q\&A data: Time series of all events (background) and new user
  arrivals(front curve) with the time interval 10 minutes. Bottom: Zipf’s and Heap’s law plot
  for all tags (purple) and all distinct combinations of tags---innovation
  (black).}
\label{fig-TS-Zipf-QA}
\end{figure}

An example of the time series of the number of events in the Q\&A data is shown in Fig.\ \ref{fig-TS-Zipf-QA}, top panel. We also show the time series of the number of new users, who bring further questions and thus drive the system's dynamics. It should be noted that the occurrence of the power-laws as the signature of the mechanisms of SOC  in these
systems \cite{we-SOC2017}, is also compatible with the power-law in the ranking distribution (Zipf's law) and with it related Heap's law. Here in the lower panels in Fig.\ \ref{fig-TS-Zipf-QA}, we show the corresponding plots for the frequency of tags and for the appearance of unique combinations of tags, which describes the innovation expansion in this knowledge creation process. A more detailed study can be found in \cite{we-SciRep2015}.

\begin{figure}[!htb]
\begin{tabular}{cc} 
\resizebox{16pc}{!}{\includegraphics{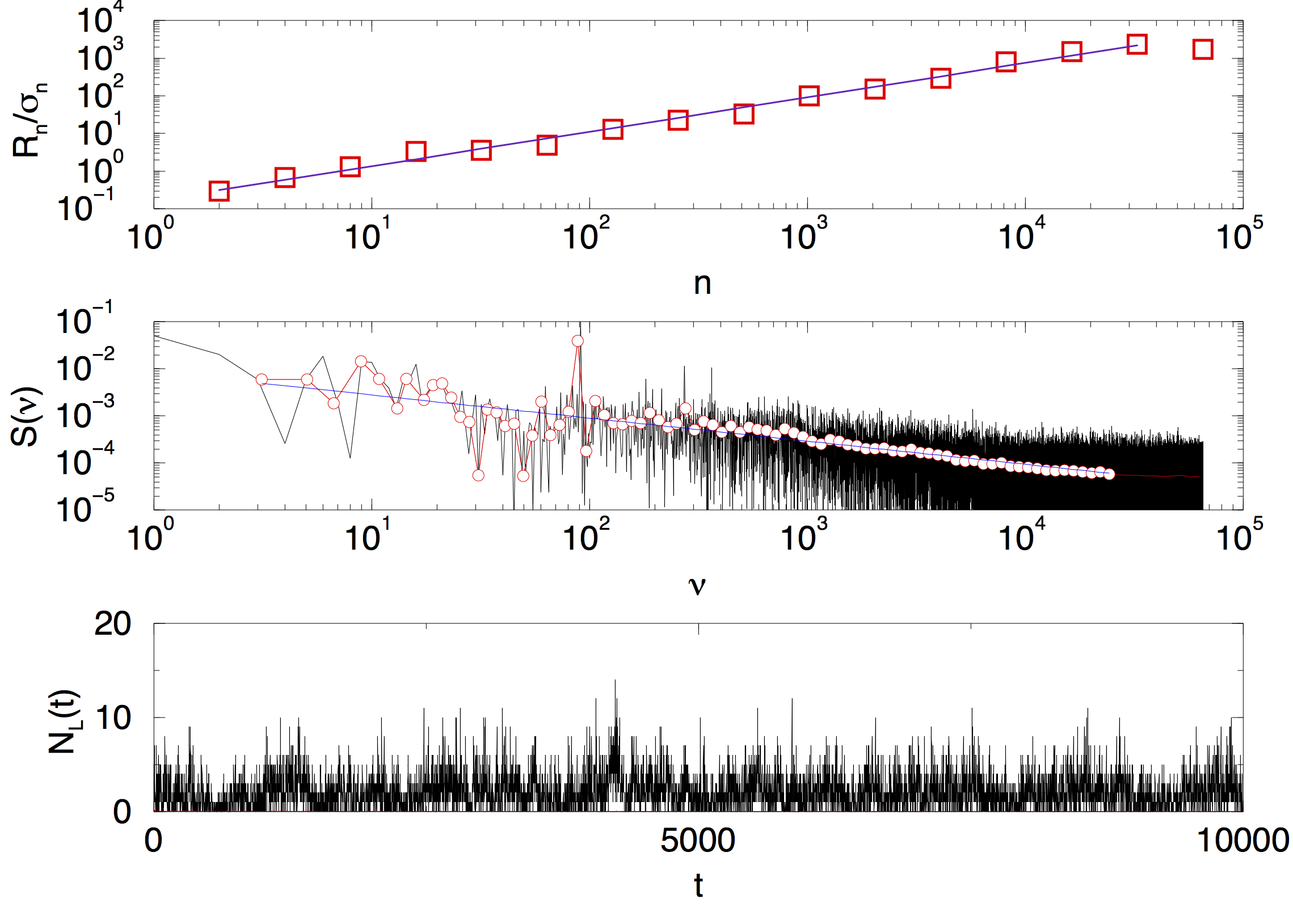}}\\
\end{tabular}
\caption{From Ubuntu chats data: Time series of the number of new
  links appearing in a small time window t, bottom panel, and its
  power spectrum, middle, and the fluctuations defining the Hurst
  exponent, top panel.}
\label{fig-SOC-chats}
\end{figure}
\begin{figure}[!htb]
\begin{tabular}{cc} 
\resizebox{20pc}{!}{\includegraphics{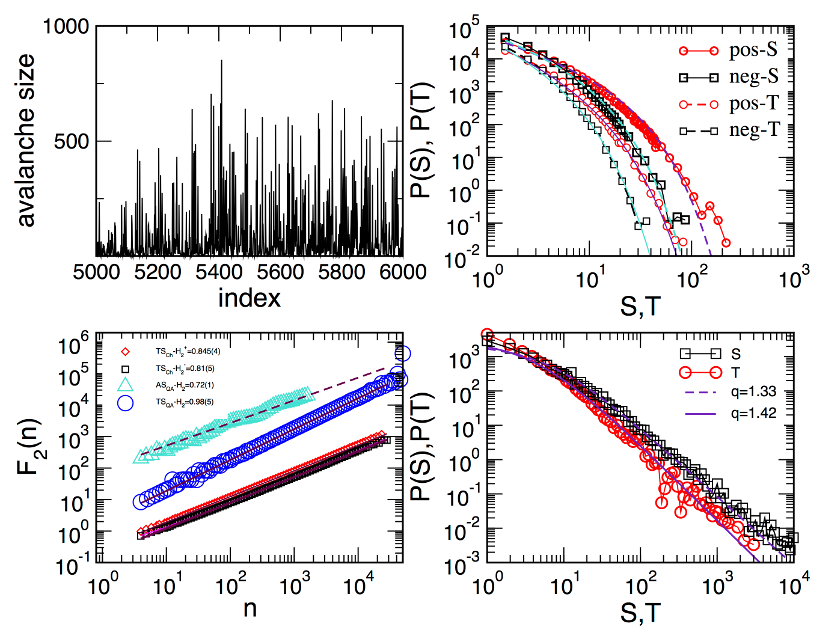}}\\
\end{tabular}
\caption{ Top-left: Avalanche sequences in Q\&A; Bottom-left: Scaling
  of the standard deviation function for time-series (TS) and
  avalanche series(AS) in Chats and  Q\&A data; the corresponding Hurst
  exponents are shown in the legend. Top-right:  Avalanche
  distributions $P(S)$ and $(T)$ for chats with positive and negative
  emotion contents; fits according to stretch-exponential distribution
 $P(X)=aX^{-\tau_X}\exp(-(X/b)^{\sigma_X})$ with $\tau_S\sim 1$ and
 $\tau_T\sim 1.25$ and stretching $\sigma_S\sim 0.9$,  $\sigma_T\sim
 0.85$ both for positive and negative contents.  Bottom-right: For
 Q\&A data, the
 distributions of the avalanche sizes and durations, fitted by
 $q$-exponential with the $q$ values in the legend.  }
\label{fig-SOC-comb}
\end{figure}

In the case of Ubuntu chats, which results in the emotion-based social
linking, the self-organising dynamics can be detected in the time
series of the appearance of new links, as shown  Fig.\ \ref{fig-SOC-chats}.
Specifically,  the long-range temporal correlations of the appearance
of new links manifest themselves
in the power-law decay of the power spectrum according to
\begin{equation}
S(\nu) \sim 1/\nu^{\phi} \;
\label{eq-PS}
\end{equation}
where $\phi\sim 1$XX. Moreover, the number of new
links fluctuates over time according to the fractal (Hurst) exponent
$H_2\sim 1$. 
By definition, this exponent is related to the standard deviations of the
fluctuations around a local trend $y_\mu(t)$ of the integrated signal
$Y(i)=\sum_{k=1}^i(N_c(k)-\langle N_c\rangle)$ on the $\mu$-th segment
of the length $n$,
i.e.,
\begin{equation}
F_2(n) = \left[\frac{1}{N}\sum_{\mu=1}^{N}F^2(\mu,n)\right]^{1/2}
\sim n^{H_2}\;
\label{eq-F2}
\end{equation}
where $F^2(\mu,n)= (1/n)\sum_{i=1}^n[Y((\mu-1)n+1)-y_\mu(i)]^2$.
Thus, by varying the length of the segment $n$, the fluctuation
function is determined and plotted against $n$. The slope of this plot
then defines the Hurst exponent, see more plots in Fig.\ \ref{fig-SOC-comb}.

In the SOC states, the size $s$ and duration $T$ of avalanches (cf. caption
to Fig.\ \ref{fig-TS-example} for the precise definition) are expected
to have a power-law decay. We determine these distributions directly from the time series of the number of events in
Q\&A data and from the chats data in the presence of a Bot which favours positive/negative emotions. The results that are shown in Fig.\ \ref{fig-SOC-comb} indicate that the avalanches have a substantial scale invariance in both knowledge-sharing processes. However, we find that different mathematical expressions fit the probability distributions of the avalanches. Explicitly, in the chat data, the power-law decay in a short interval is followed by an
exponential cut-off. The exponents and the cut-offs are slightly
different for the layer with positive/negative emotion flow. 
In the case of data Q\&A data, however, the avalanches obey a
$q$-exponential distribution 
\begin{equation}
P(X) =B_X\left[1-(1-q)\frac{X}{X_0}\right]^{1/1-q} \;
\label{eq-qexp}
\end{equation}
where the exponent $q\neq 1$ is recognised as the nonextensivity parameter
\cite{tsallis,pavlos}. Moreover, the sequence of avalanches, see the
top-left panel  in Fig.\ \ref{fig-SOC-comb}, was shown to have
a multifractal spectrum \cite{we-SOC2017}.

\section{Discussion and Conclusions\label{sec:discussion}}

Knowledge sharing on online social sites represents a particular type of social dynamics resulting in the emergent networks of a characteristic structure. Considering two types of online social interactions which involve knowledge sharing, we have shown that different sorts of networks emerging from these interactions, in particular, the explicit-knowledge networks built via Q\&A and the social graph built by Ubuntu chats,  are $\delta$-hyperbolic. Furthermore, they 
exhibit a rather low $< \delta > \lesssim 0.33$ and  $\delta_{max}$ that
does not exceed 2.  Our analysis
\cite{we-SciRep2015,we-KCNets_PLOS2016,we-Chatnets,we-MySpace11,we-MyspaceAT}
shows that the occurrence of the higher organised structure with simplicial complexes in these networks can be regarded as the origin of their hyperbolic geometry.  
 The networks underlying knowledge creation via Q\&A occur in the stochastic
process with the prominent signatures of self-organised criticality
\cite{we-SOC2017}.  In the case of Ubuntu chats, the temporal correlations, fractality and avalanches of events are
also apparent; however, the avalanche distributions indicate possible parameter dependence and a dynamical phase transition, rather than a SOC  attractor. These global states emerge from the different use of knowledge contents at the
elementary scale of human interactions. This question requires
additional study.

The growth of the collective knowledge appears over time by building
innovation channels on the network of used knowledge contents. Their structure can be decomposed into cliques of all orders up to an existing $q_{max}$. Having in mind that these elementary geometric
descriptors of the network are $\delta$-hyperbolic \cite{HB1,gromov-4p} with $\delta$=0, it is
conceivable that their complexes, which make the graph, also exhibit hyperbolic structure with a low value of $\delta$, as shown in Fig.\
\ref{fig-Chats-HB}. Note that architecture with simplicial complexes is also found in the social network MySpace \cite{we-MyspaceAT}. These results open a new direction of research of the origin of hyperbolicity in the online social graphs.
The network of Ubuntu chats, on the other hand, grows around an active core of knowledgeable users and Bot by attaching new users, who then stay to serve new arrivals further. In this way, based on the experience of users, a hierarchical architecture of the network is built over time. The characteristic $k$-core structure of this whole network, as well as its layers carrying emotional messages \cite{we-Chatnets}, is
compatible with the observed hyperbolicity in Fig.\ \ref{fig-Chats-HB}.

Although the considered networks are of a different nature, in both cases, the knowledge (expertise) of the participating individuals plays an essential role in building the system at each stage. More precisely, in each interaction, specific knowledge content is required to meet the current needs. This knowledge-matching constraints provide a delicate balance at the elementary scale and, consequently, implies the logical attachment of the primary forms into an extensive geometry, eventually resulting in a hyperbolic structure.  Note that, such a delicate balance which is apparent in knowledge-sharing processes is not present in various other social interactions; nevertheless,  they can result in hyperbolic networks, which are based on another optimisation principle. 

\section*{\normalsize Acknowledgments}
This work is based on the results of several previous publications,
for which I thank the collaborations with Milovan \v{S}uvakov, Miroslav
Andjelkovi\'c, Marija Mitrovi\'c Dankulov, Vladimir Gligorijevi\'c, Milan
Rajkovi\'c, and Roderick Melnik. Work supported  by  the Slovenian
Research Agency (research code funding number P1-0044).

\end{document}